\documentclass[aip,sd,amsmath,amssymb,reprint]{revtex4-1}

\usepackage{graphicx}
\usepackage{dcolumn}
\usepackage{bm}
\usepackage{epstopdf}
\usepackage{xcolor}
\usepackage{natbib}
\usepackage{subfigure}

\newcommand{\wi}[1]{^{({\rm #1})}}
\newcommand{\wLH}{\wi{lh}}
\newcommand{\wB}{\wi{ebw}}
\newcommand{\wX}{\wi{x}}
\newcommand{\wQ}{^{(q)}}
\newcommand{\Eq}[1]{Eq.~(\ref{#1})}

\newcommand{\rv}[1]{\textcolor{black}{#1}}

\begin{document}

\title{Kinetic simulations of X-B and O-X-B mode conversion {and its deterioration at high input power}}

\author{A. V. Arefiev}
\affiliation{Institute for Fusion Studies, The University of Texas, Austin, Texas 78712, USA}
\affiliation{Center for High Energy Density Science, The University of Texas, Austin, Texas 78712, USA}

\author{I. Y. Dodin}
\affiliation{Princeton Plasma Physics Laboratory, Princeton University, Princeton, New Jersey 08543, USA}

\author{A. K\"ohn}
\affiliation{Max Planck Institute for Plasma Physics, D-85748 Garching, Germany} 

\author{E. J. Du Toit}
\affiliation{York Plasma Institute, Department of Physics, University of York, York YO10 5DD, UK}

\author{E. Holzhauer}
\affiliation{Institute of Interfacial Process Engineering and Plasma Technology, University of Stuttgart, D-70569 Stuttgart, Germany}

\author{V. F. Shevchenko}
\affiliation{{Tokamak Energy Ltd, 120A Olympic Avenue, Milton Park, Abingdon OX14 4SA, UK}}

\author{R. G. L. Vann}
\affiliation{York Plasma Institute, Department of Physics, University of York, York YO10 5DD, UK}

\date{\today}

\begin{abstract}
{Spherical tokamak plasmas are {typically} overdense and thus inaccessible to {externally-injected} microwaves in the electron cyclotron range. The electrostatic electron Bernstein wave (EBW), however, provides a method to access the plasma {core} for heating and diagnostic purposes. Understanding the details of the coupling process to electromagnetic waves is thus important both for the interpretation of microwave diagnostic data and for assessing the feasibility of EBW heating and current drive. While the coupling is {reasonably} well{--}understood in the linear regime, {nonlinear physics arising from high input power has not been previously quantified}. To tackle this problem, we have performed one- and two-dimensional fully kinetic particle-in-cell simulations of the two possible coupling mechanisms, namely X-B and O-X-B mode conversion. We find that the ion dynamics has a profound effect on the field structure in the nonlinear regime, as high amplitude short-scale oscillations of the longitudinal electric field are excited in the region below the high-density cut-off prior to the arrival of the EBW. We identify this effect as the instability of the X wave with respect to resonant scattering into an EBW and a lower-hybrid wave. We calculate the instability rate analytically and find this basic theory to be in reasonable agreement with our simulation results.}
\end{abstract}

\maketitle


\section{Introduction}

In spherical tokamaks the plasma frequency usually exceeds the electron cyclotron frequency (ECF) over the whole confinement region, characterizing it as overdense. Electromagnetic waves in the ECF range can therefore not be used \rv{directly} for heating or current drive\rv{. This is a} particular disadvantage for spherical tokamaks as they rely on efficient non-inductive current drive mechanisms due to only very little space for a central transformer coil. Electron Bernstein waves (EBWs)~\cite{Laqua2007} are electrostatic waves which can be used instead, since they are very well absorbed at the ECF and its harmonics and provide an efficient method for driving toroidal net currents~\cite{Urban2011}. Understanding the details of the coupling process between electromagnetic waves and EBWs is important for assessing feasibility studies of EBW heating and current drive and also for interpreting diagnostics involving EBWs.

At low to intermediate power levels, excitation of EBWs and subsequent current drive has been demonstrated successfully in spherical tokamaks~\cite{Shiraiwa2006, Shevchenko2015}. The linear regime has been studied numerically with full-wave codes~\cite{Kohn2014} and is well understood. On the other hand, the EBW physics in the nonlinear regime, which is of interest in the context of heating and current drive in modern fusion devices, still presents a challenge~\cite{Xiang2011, Asgarian2014}. A complete analytical description of the nonlinear regime is not generally feasible and therefore numerical simulations are required.

A self-consistent description of electron kinetics and wave propagation must resolve spatial and temporal scales associated with the electron gyro-motion in order to correctly recover the physics relevant to EBWs. One suitable approach is the particle-in-cell (PIC) framework that uses macro-particles to simulate electrons and ions kinetically. In a recent publication~\cite{Arefiev2015}, we reported results of one-dimensional (1D) and two-dimensional (2D) simulations performed in a linear regime using \rv{the} particle-in-cell code EPOCH~\cite{EPOCH}. (For another relevant \rv{project} involving PIC simulations of upper-hybrid (UH) waves see Ref.~\cite{Xiao2015}) These simulations successfully reproduce X-B and O-X-B mode conversion, with both the initial electromagnetic mode and the excited EBW matching the linear dispersion relations. These benchmarking runs confirm that PIC codes are suitable for studies of EBW excitation and, most importantly, they identify the simulation parameters needed to correctly reproduce the mode conversion. 

In this work, we use 1D and 2D PIC simulations to examine the mode conversion and associated physics in nonlinear regimes where the wave amplitude is sufficiently high to appreciably affect the thermal electron motion. We find that the ion mobility has a profound effect on the wave electric field structure in the nonlinear regime. Our kinetic simulations show that high-amplitude short-scale oscillations of the longitudinal electric field are excited in the region below the high-density cut-off prior to the arrival of the EBW generated at the upper-hybrid resonance via a mode conversion. Simulations performed with immobile ions show no such short-scale oscillations. Using a detailed analysis of the ion dynamics in the simulations, we have identified the plasma oscillations driven in the nonlinear regime as lower-hybrid (LH) oscillations. This is in agreement with earlier studies~\cite{Gusakov2007, McDermott1982, Porkolab1982, Lin1981} where the X-wave was reported to become unstable with respect to backscattering into an EBW and an LH wave. Therefore, our present work can be understood as the first study where this instability is observed in first-principle PIC simulations. We also compare our numerical results with the theory of this scattering instability. However, an in-depth theoretical analysis is left to a companion paper~\cite{Dodin2017}. 

The rest of the paper is organized as follows. In Sec.~\ref{Sec-setup}, we describe the setups that we use in our PIC simulations to examine the EBW excitation in the nonlinear regime. Sections~\ref{Sec_X-B}  and \ref{Sec-Ion-Dynamics} present simulations of the X-B conversion in the nonlinear regime and the corresponding analysis of the ion dynamics. Simulations of the O-X-B conversion in the nonlinear regime are discussed in Sec.~\ref{Sec-OXB}. Section~\ref{Sec-Theory} provides a theoretical model that we use to explain the origin of the observed instability. Finally, in Sec.~\ref{Sec-Summary}, we summarize our results and discuss their implication for future experiments with high input power.

\begin{figure*} 
\centering
\includegraphics[width=1.8\columnwidth]{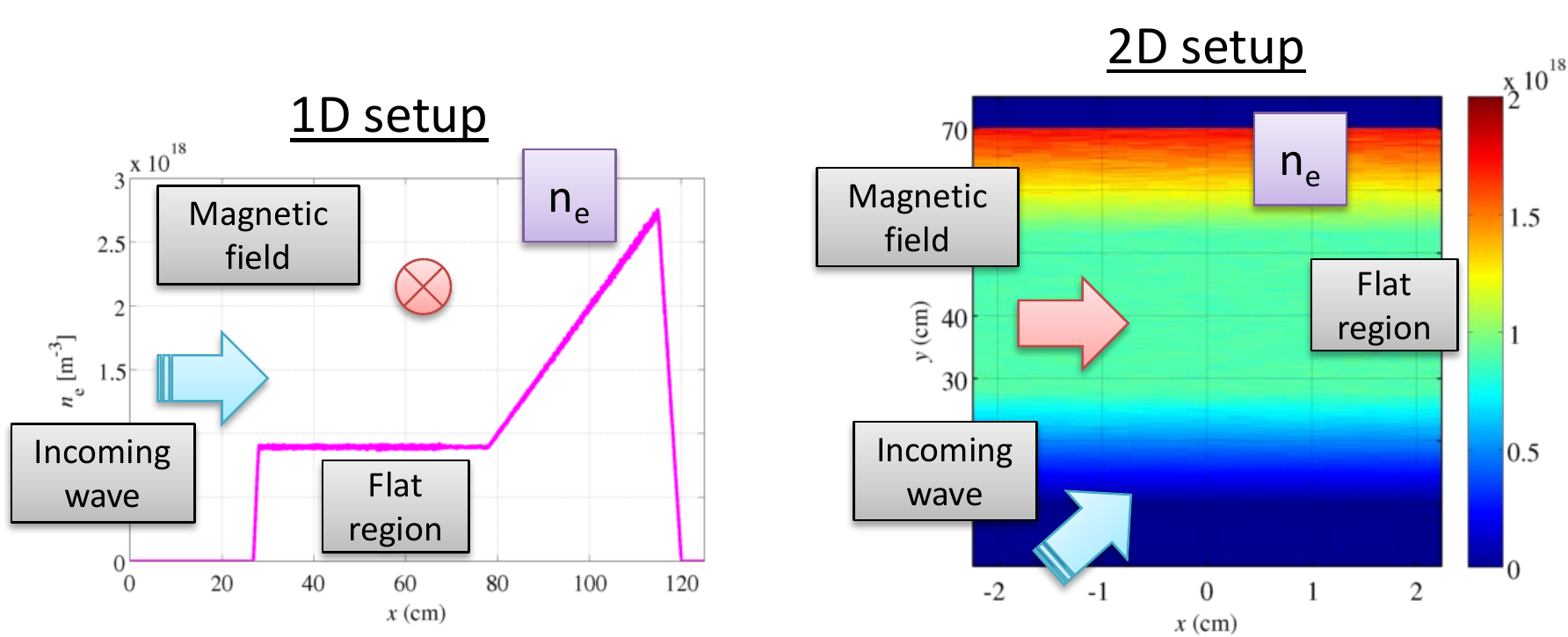}
\caption{One and two-dimensional setups for PIC simulations of the X-B and O-X-B mode conversion~\cite{Arefiev2015}.} \label{Fig1}
\end{figure*}


\section{1D and 2D PIC setups for studying EBW excitation} \label{Sec-setup}

In a recent publication~\cite{Arefiev2015}, we proposed 1D and 2D setups that are suitable for simulation and clear identification of the EBW excitation using a standard PIC code (Fig.~\ref{Fig1}). Both setups have common features that need to be emphasized. A plane incoming wave is injected into a vacuum gap that is deliberately introduced between the plasma and the computational domain boundary on the injection side. A uniform magnetic field $B_0$ that is parallel to the injection boundary is initialized throughout the simulation domain. ($B_0$ is directed along the $y$-axis in the 1D setup and along the $x$-axis in the 2D setup.) The plasma density profile has three distinct regions. Region I, which is on the injection side, corresponds to a steep density profile that includes the UH resonance (UHR). Region II, which is in the middle, corresponds to an extended flat density region. Region III, which is on the opposite side, corresponds to a steep density profile that includes a high-density cut-off for an X-mode.

The multi-gradient density profile makes it convenient to both excite and observe EBWs in the X-B and O-X-B conversions in simulations. Region I is where the EBW excitation takes place. The density gradient in this region needs to be very steep in order to simulate the X-B mode conversion. This scenario relies on X-mode tunneling past the low-density cut-off and subsequent excitation of an EBW in the vicinity of the \rv{UH} layer. The efficiency strongly depends on the distance between the low-density cut-off and the \rv{UH} layer and, therefore, it can be dramatically increased by using a steep density gradient. Region II is introduced to aid EBW identification following its excitation. There is no density gradient in this extended region, so that the dispersion relation of propagating waves remains unchanged. Finally, Region III is introduced to provide a high-density cut-off for an X-mode. The cut-off reflects an X-mode propagating from Region II up the density gradient. We use the steepness of the density profile in this region primarily to minimize the size of the simulation domain. There is a vacuum gap between this region and the boundary of the simulation domain in order to eliminate possible numerical boundary artifacts in particle dynamics. 

The 2D setup (right panel of Fig.~\ref{Fig1}) allows for the incoming waves to be injected at an angle to the applied magnetic field (directed along the $x$-axis) and to the density gradient (directed along the $y$-axis). This setup is therefore helpful for achieving and observing an O-X-B mode conversion. Even though a full 2D simulation in which an incoming wave packet has a finite transverse width is possible, it is computationally demanding because the box has to be sufficiently wide to accommodate the entire path of the wave packet. A ``reduced'' 2D setup is needed to facilitate multiple runs that are necessary for parameter scans.  Such a reduced setup is achieved by using periodic boundary conditions (for particles and fields) on the left and right side of the domain shown in Fig.~\ref{Fig1} (right panel). In this setup, the injected wave is a plane wave with given wave-vector components parallel ($k_{\parallel} \equiv k_x$) and perpendicular ($k_{\perp} \equiv k_y$) to the magnetic field. The transverse size of the domain (along the $x$-axis) is set to $2 \pi / k_{\parallel}$ to account for the wave periodicity. We use open boundary conditions at the top and bottom boundaries of the simulation domain. Note that the 1D setup similarly employs open boundary conditions.

\begin{figure*} 
 \centering
  \includegraphics[width=1.5\columnwidth]{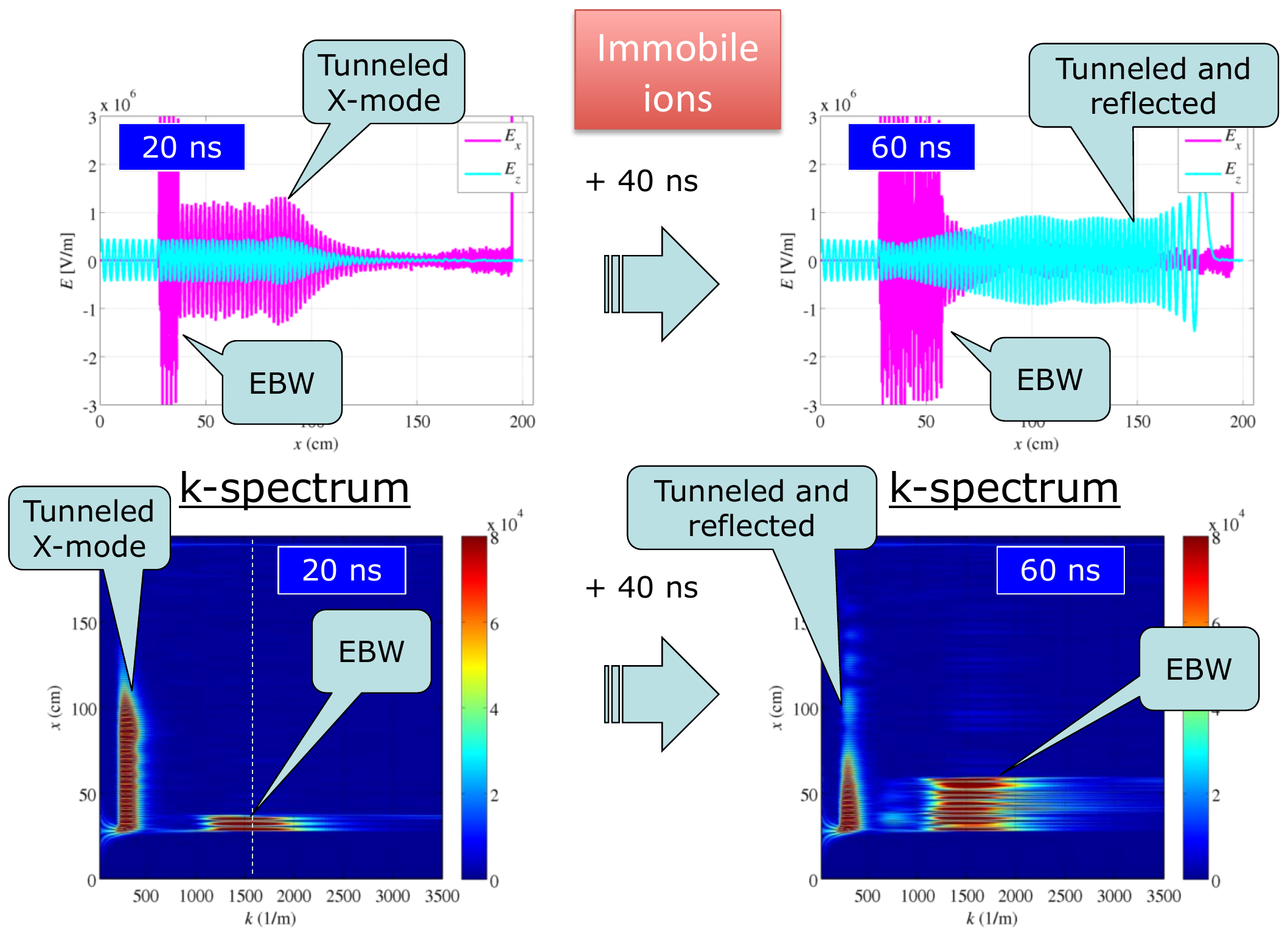}
  \caption{1D PIC simulation of the X-B mode conversion in a nonlinear regime with \textit{immobile} ions.}
  \label{Fig2}
\end{figure*}

Ultimately, the following considerations should be kept in mind when choosing the density in the flat region in both setups: 1) if O and X modes are not desired in the flat region (Region II), then its density should be above the high-density cutoff; 2) the wavelength of the EBW that decreases with density must be large enough compared to the grid size in order to resolve the EBW oscillations; 3) the amplitude of the numerical field fluctuations that increases with density for a fixed number of macro-particles per cell must be much less than the amplitude of the expected EBW signal.

Using the described 1D and 2D setups, we have performed PIC simulations of EBW excitation by waves with a relatively low amplitude ($10^5$ and $5 \times 10^4$ V/m). We have reproduced X-B and O-X-B mode conversion, with the O-mode, X-mode, and the excited EBW matching their linear dispersion relations~\cite{Arefiev2015}. These benchmarking runs confirm that \rv{PIC simulation} is suitable for studies of EBW excitation and, most importantly, they identify the simulation parameters needed to correctly reproduce the mode conversion. In this work, we use the 1D and 2D setups to examine the mode conversion in nonlinear regimes, where the wave amplitude is sufficiently high to affect the thermal electron motion.

\begin{figure*} 
\centering
  \includegraphics[width=1.5\columnwidth]{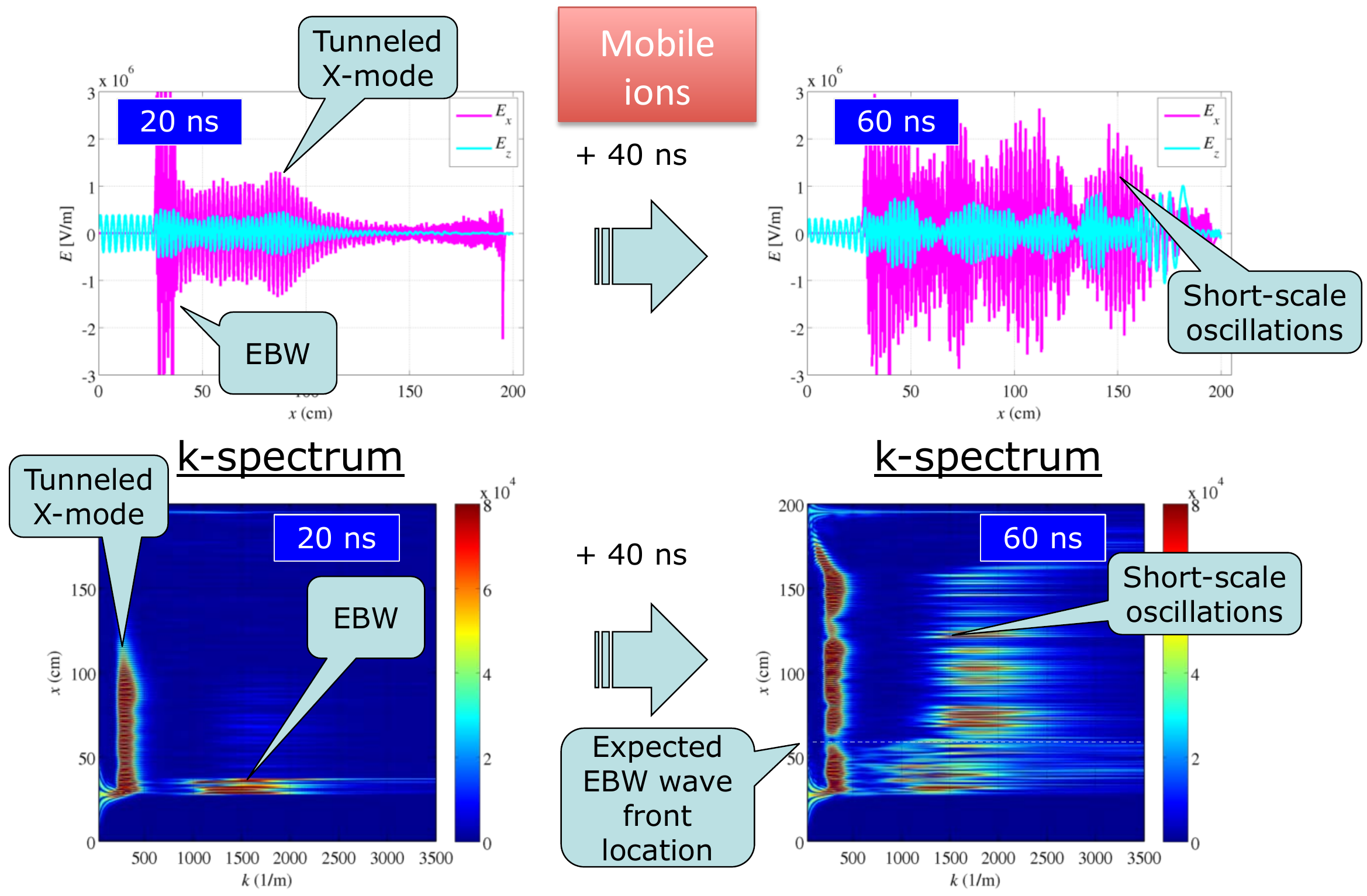}
  \caption{1D PIC simulation of the X-B mode conversion in a nonlinear regime with \textit{mobile} ions.}
  \label{Fig3}
\end{figure*}


\section{X-B conversion in a nonlinear regime} \label{Sec_X-B}

The primary objective of our study is to investigate excitation of EBWs in a nonlinear regime that we access by increasing the amplitude of the incoming wave. In this regime, the wave distorts the thermal motion of electrons. A PIC code simulation is a fully-kinetic simulation that resolves the electron dynamics, which makes it a well-suited approach for this problem. In the present work we use the MPI-parallelized explicit second-order relativistic code EPOCH~\cite{EPOCH}.

We \rv{first study} an X-B mode conversion \rv{scenario} that can be examined in a 1D simulation. In the simulations presented below, we use a confining magnetic field $B_0 =  0.25$ T and an incoming wave with frequency $f = 10$ GHz. The peak amplitude of the incoming wave that follows a gradual initial ramp-up is set to $8 \times 10^5$ V/m. The size of the computational domain is 2 m ($6 \times 10^4$ cells). The wave is injected at the left boundary located at $x = 0$. The incoming wave is an X-mode whose electric field is polarized along the $z$-axis.

The initial electron density profile is adopted in the form
\begin{equation}
\left. n_e \right/ n_{crit} =
   \begin{cases}
      0,  & \text{for $x <  x_{b1}$;}\\
      128(x-x_{b1})/l,  & \text{for $x_{b1} \leq x <  x_L$;}\\
      0.72,  & \text{for $x_L \leq x \leq  x_R$;}\\
      4 (x - x_2)/l,  & \text{for $x_R < x <  x_{b2}$;}\\
      0, & \text{for $x \geq x_{b2}$,}
   \end{cases} \label{n_sim}
\end{equation}
where $n_{crit} \approx 1.24 \times 10^{18}$ m$^{-3}$ is the critical density for the \rv{chosen} frequency. We have introduced the following quantities to parameterize the multi-gradient density profile: $x_L = 0.28$ m, $x_R = 1.58$ m, $\Delta x = 0.005625$ m, $x_{b1} = x_L - \Delta x$, $x_{b2} = 1.95$ m, $x_2 = 1.4$ m, and $l = 1$ m. The electron population is initialized using 400 macro-particles per cell. The initial electron temperature is set at $T_e = 950$ eV. We also initialize deuterium ions using 400 macro-particles per cell with the ion \rv{number} density equal to $n_e$ and the ion temperature equal to $T_e$.

\begin{figure}[b] 
\centering
\includegraphics[width=0.9\columnwidth]{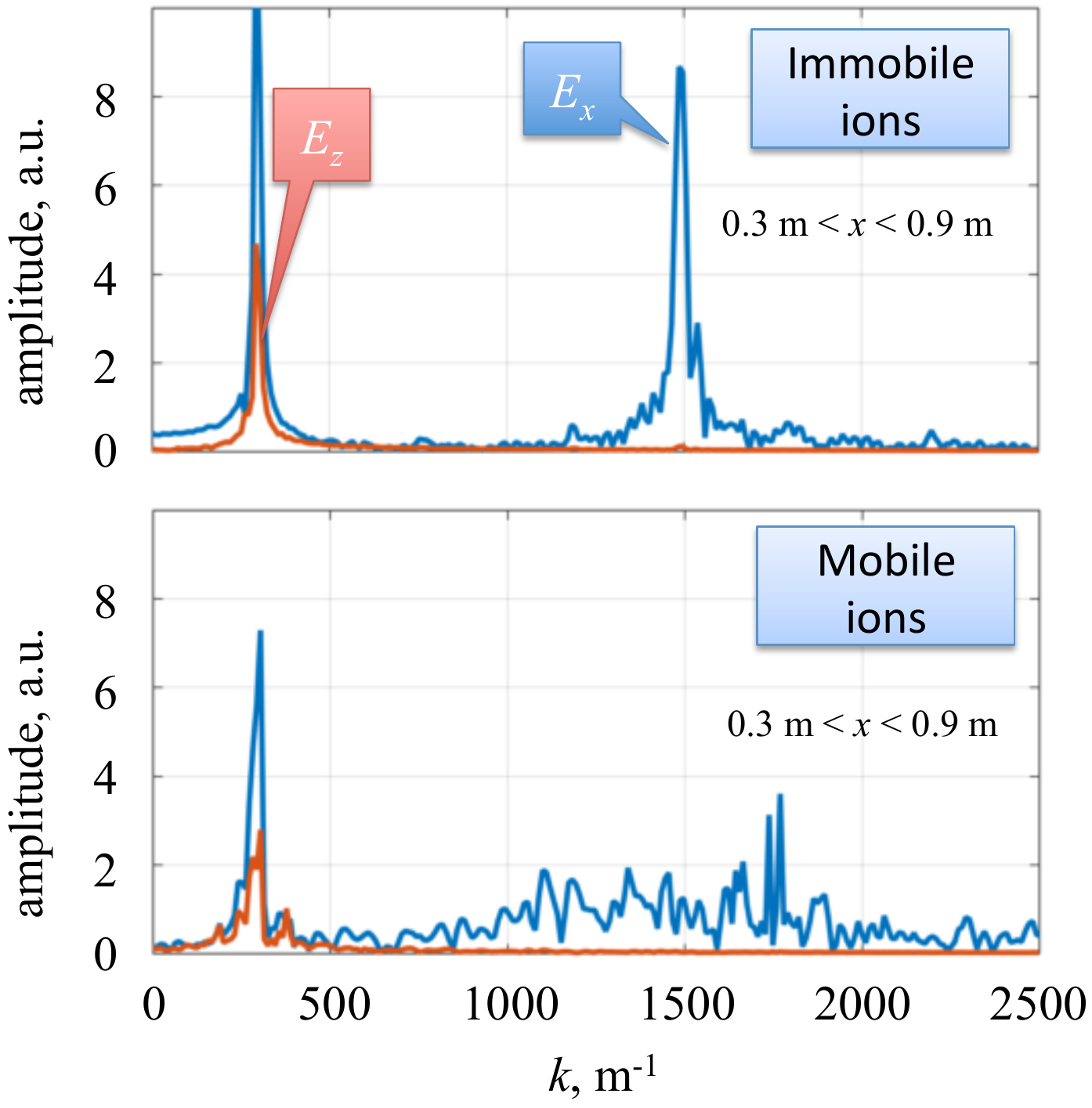}
\caption{Spectra of the transverse and longitudinal electric field components at 45 ns into the simulation in the flat region for 0.3 m $< x < $ 0.9 m.}
\label{Fig_1D_field_v1}
\end{figure}

\begin{figure}[b] 
\centering
\includegraphics[width=0.9\columnwidth]{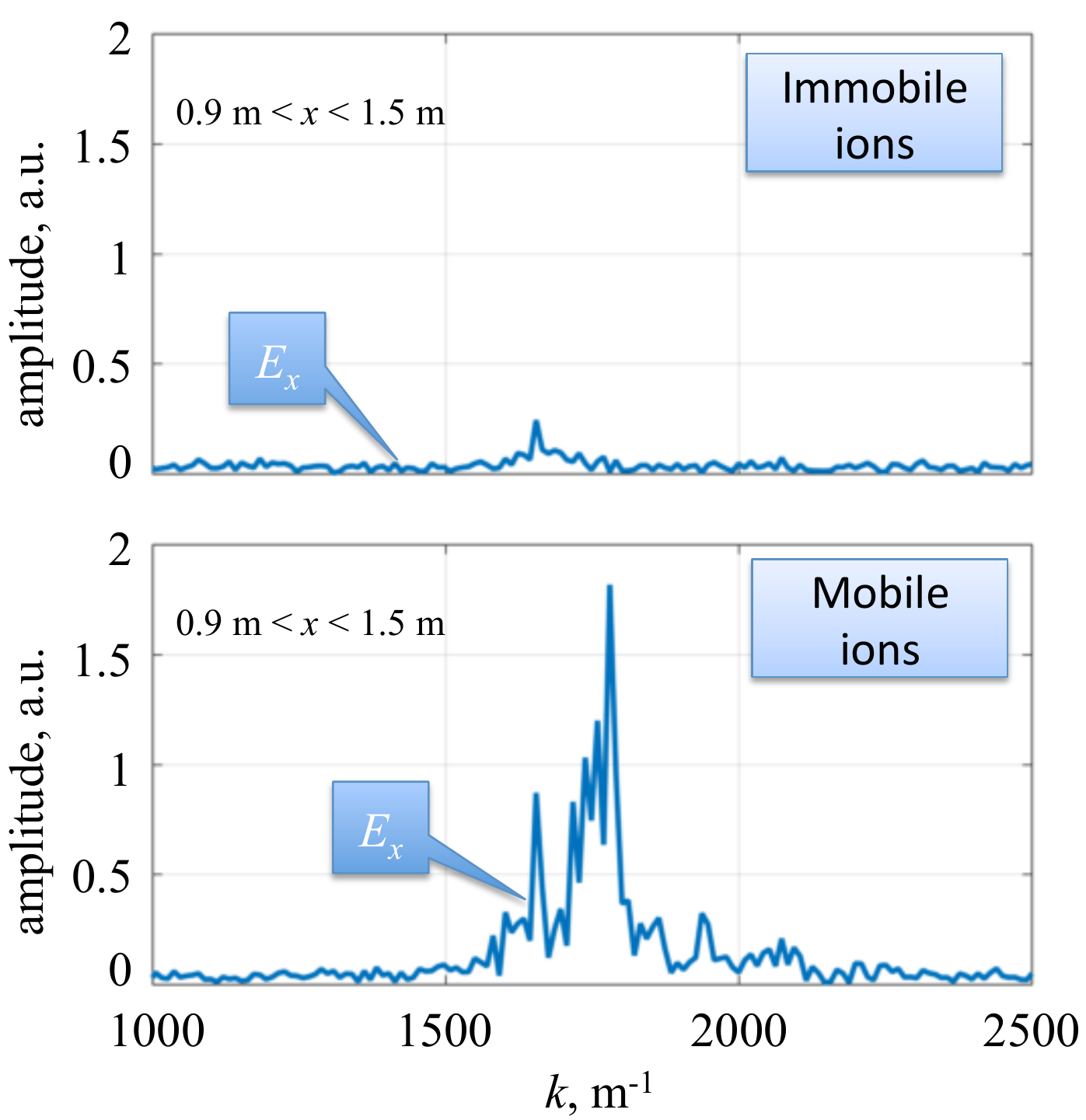}
\caption{Spectra of the longitudinal electric field at 45 ns into the simulation in the flat region for 0.9 m $< x < $ 1.5 m.}
\label{Fig_1D_field_v2}
\end{figure}

Our first simulation in the nonlinear regime was performed with immobile ions. This was motivated by the fact that the ion mobility has no significant impact on the dispersion relation of EBWs in a linear regime. Snapshots of transverse ($E_z$) and longitudinal ($E_x$) electric field components are shown in the upper panels of Fig. \ref{Fig2} at 20 ns and 60 ns into the simulation. As the X-mode tunnels into the flat density region, it excites an EBW at the sharp density gradient (upper-left panel of Fig.~\ref{Fig2}). The corresponding $k$-spectrum for the longitudinal component of the electric field $E_x$ as a function of the distance into the plasma is shown in the lower-left panel of Fig.~\ref{Fig2}. The excited EBW follows the dispersion relation predicted for the linear regime \rv{(}marked with a dashed vertical line\rv{)}. The right two panels in Fig.~\ref{Fig2} show the field profile and the $k$-space 40 ns later. By this point, the tunneled X-mode has already reflected off the high-density cut-off on the right of the simulation domain and it is moving towards the UH \rv{layer}. The basic features of the EBW spectrum however remain unaffected by this large-amplitude reflected wave, similarly to what one would expect in the linear regime.

\begin{figure*} [ht]
\centering
\includegraphics[width=1.7\columnwidth]{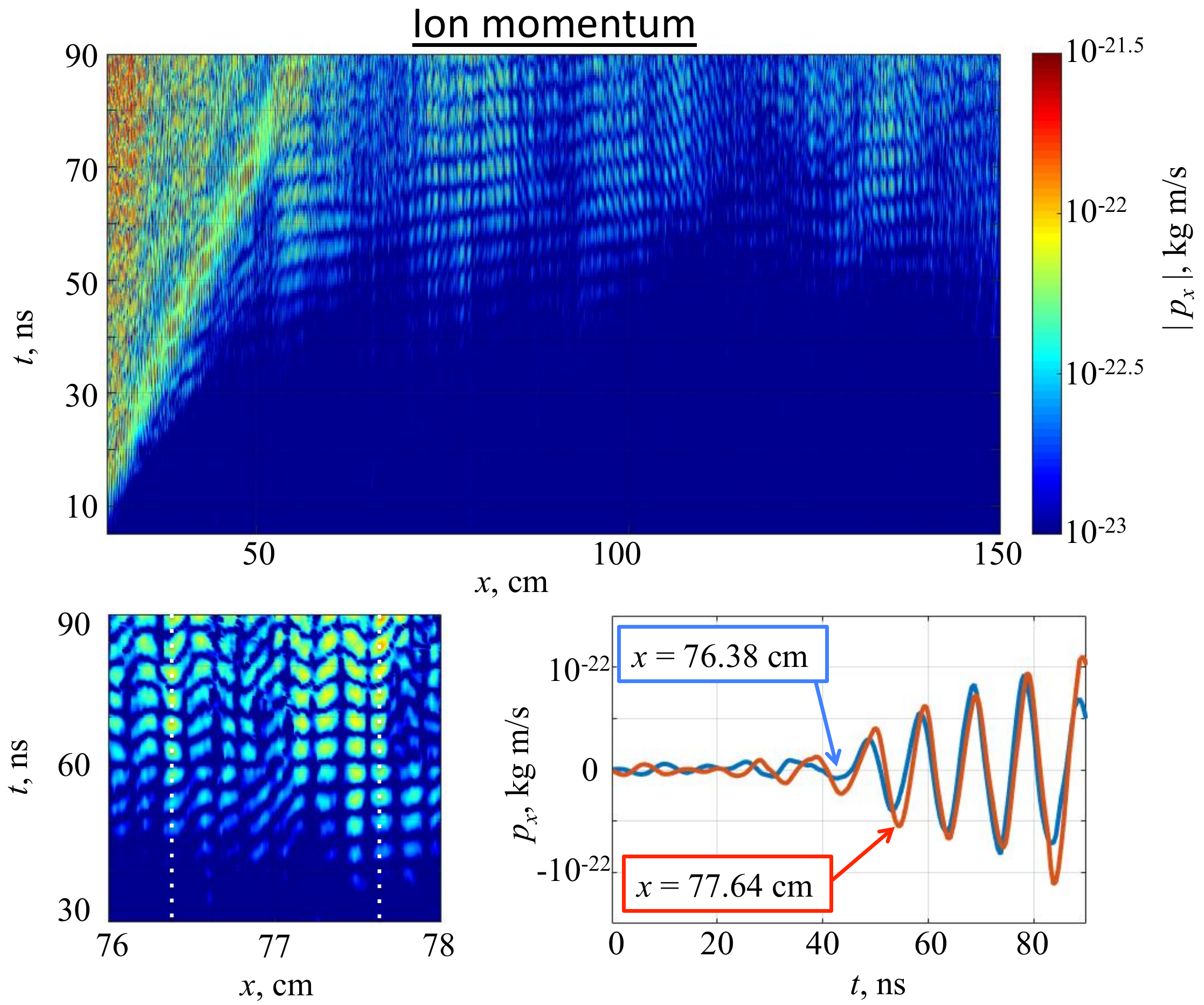}
\caption{Ion momentum in a 1D PIC simulation of the nonlinear regime with initially cold ions. The lower-left panel is a zoom-in of the upper panel for 76 cm $< x <$ 78 cm and 30 ns $< t < $ 90 ns. The lower-right panel gives a time evolution of the ion momentum at two fixed locations, $x = 76.38$ cm and $x = 77.64$ cm, marked in the lower-left panel with the dotted lines.}
\label{Fig_I_P_combo}
\end{figure*}

The second simulation in the nonlinear regime was performed with mobile ions and \rv{the} results are shown in Fig.~\ref{Fig3}.  The initial stage prior to the X-mode being reflected off the high-density cut-off (left two panels in Fig.~\ref{Fig3}) is similar to what we observed in the simulation with immobile ions (left two panels in Fig.~\ref{Fig2}). However, the field structure dramatically changes 40 ns later, as the reflected X-mode starts to propagate through the plasma. As evident from the two  right panels in Fig.~\ref{Fig3}, short-scale oscillations appear throughout the plasma. 


In order to closely examine the differences in the electric field structure for the cases of mobile and immobile ions, we have performed a Fourier analysis of the field in the flat density region at 45 ns into the simulation. It is worth noting that an EBW that is excited in a linear regime at the sharp density gradient would cover less than half of the flat region by this time. We therefore separately consider the left (0.3 m $< x <$ 0.9 m) and the right (0.9 m $< x <$ 1.5 m) side of the flat density region, with the corresponding spectra shown in Figs.~\ref{Fig_1D_field_v1} and~\ref{Fig_1D_field_v2}. In the case of immobile ions, short-scale oscillations of the longitudinal electric field are concentrated on the left side of the flat region (compare the two upper panels of Figs.~\ref{Fig_1D_field_v1} and~\ref{Fig_1D_field_v2}). Their spectrum has a very pronounced peak centered around $k \approx 1500$ m$^{-1}$. This is the EBW mode that is excited at the sharp density gradient and \rv{which} propagates from left to right. 

In the case of mobile ions, the short-scale part of the $E_x$ spectrum (see lower panels of Figs.~\ref{Fig_1D_field_v1} and~\ref{Fig_1D_field_v2}) has a qualitatively different structure from that in the case of immobile ions. The two striking differences are: (i) the absence of a sharp peak centered at $k \approx 1500$ m$^{-1}$ corresponding to an EBW excited in the linear regime; (ii) strong short-scale oscillations on the right side of the flat density region. It is worth emphasizing that the right side is not accessible on the considered time scale (45 ns) to an EBW propagating from the sharp density gradient on the left. We can then conclude that the ion mobility profoundly changes the interaction between different modes, as it enables excitation of short-scale electric field oscillations without the X-B mode conversion. In other words, deep inside the density plateau, the EBW is not produced through mode conversion but rather results from a local instability of the X wave.



\section{Ion dynamics in the nonlinear regime} \label{Sec-Ion-Dynamics}

In the previous section, we showed that the ion mobility has a profound impact on the electric field structure in the nonlinear regime. In what follows, we explore in detail the ion dynamics in the nonlinear regime in order to gain a better insight into the observed phenomenon. 

We have performed an additional simulation in the nonlinear regime with mobile ions, but in this case we initialized the ions as cold. A comparison of the wave structure between this simulation and the simulation with hot ions shows that the wave structure reported in Sec.~\ref{Sec_X-B} remains for the most part unaffected by the ion thermal motion or the lack thereof. We can then conclude that the observed change in the wave structure in the nonlinear regime is primarily related to the ion mobility rather than the ion temperature.

\begin{figure} 
\centering
\includegraphics[width=0.9\columnwidth]{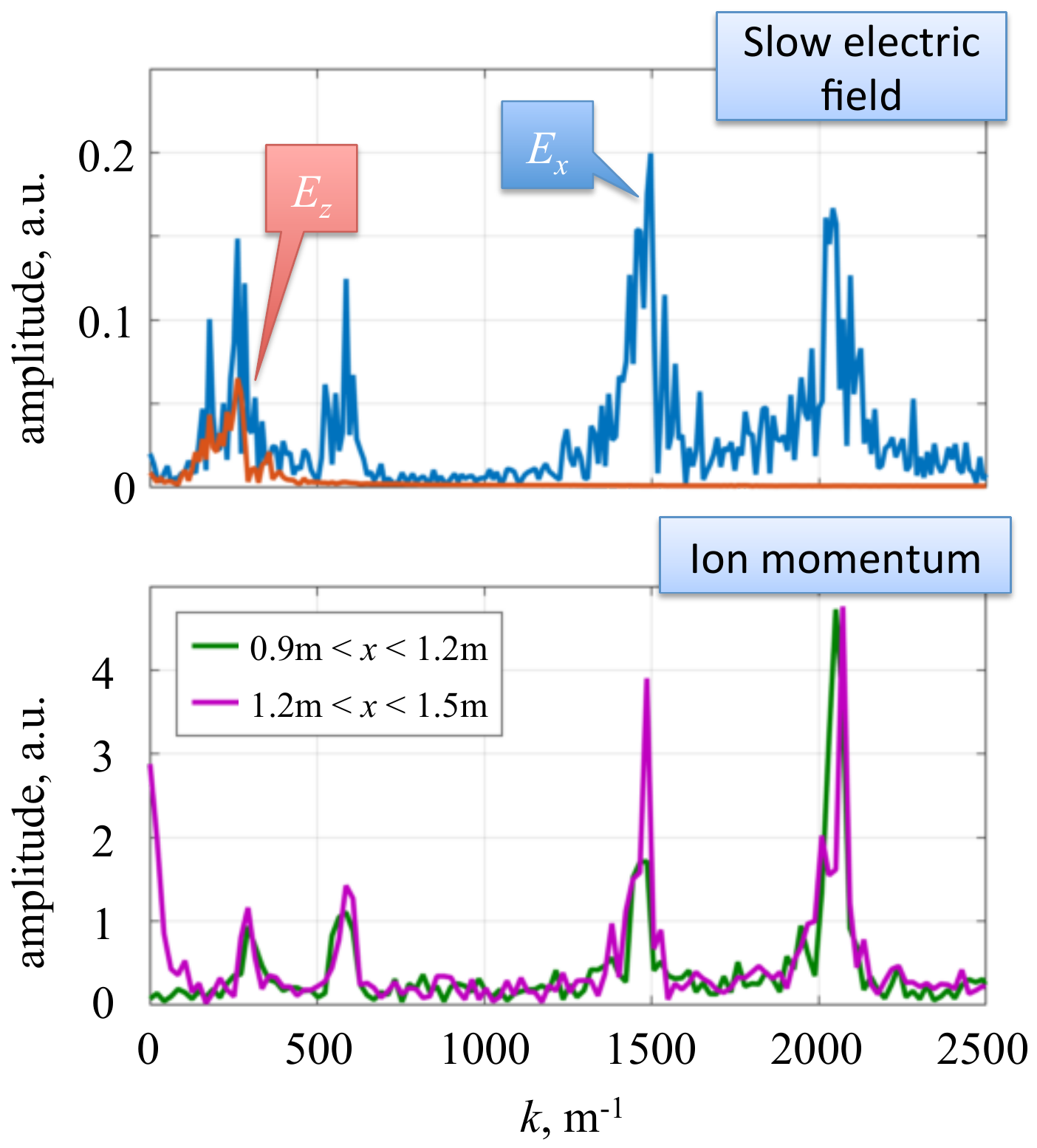}
\caption{Spectra of a slow electric field (upper panel) and of the longitudinal ion momentum (lower panel) at 59 ns into the simulation with initially cold ions. The spectra are calculated in the flat region for 0.9 m $< x < $ 1.5 m. The slow electric field is a time-average of the electric field over two periods of the incoming wave ($f = 10$ GHz).}
\label{Fig_slow}
\end{figure}

In order to examine the ion dynamics in the nonlinear regime, we take a closer look at the evolution of the ion momentum. If the ions are initially hot, then their thermal motion makes it technically challenging to decouple and visualize collective modes in the momentum space. However, since the thermal motion is not a key factor, we can examine the ion dynamics for the case of initially cold ions instead. In this case, the longitudinal ion momentum $p_x$ effectively becomes just a function of $x$ and $t$, since ions at the same location have a similar momentum. 

The corresponding plot of the ion momentum amplitude in the flat density regions is shown in the upper panel Fig.~\ref{Fig_I_P_combo}. Not surprisingly, there is a region with strong oscillations that expands from the left side of the domain. This region is associated with an EBW that is excited at the sharp density gradient that is located just to the left of the plotted region. What is surprising is that, in addition to these oscillations, there are also pronounced ion oscillations that seem to originate within the flat density region. The lower-left panel provides a zoomed-in version of the original ion momentum plot. It is evident from this plot that the ions perform periodic longitudinal oscillations whose amplitude grows in time. The lower-right panel gives a time evolution of $p_x$ at two fixed locations that are marked in the lower-left panel with the dotted lines. It is important to point out that the characteristic time scale of these oscillations is much larger than the period of the incoming wave.

The lower panel of Fig.~\ref{Fig_slow} shows the corresponding $k$-spectrum of the ion momentum oscillations in the flat density region. In order to isolate the oscillations that originate in the flat density region, the Fourier analysis was performed at $t = 59$ ns for 0.9 m $< x < $ 1.2 m and 1.2 m$< x <$ 1.5 m. It is instructive to compare the obtained 
spectrum that has two short-scale modes ($k \approx 1500$ m$^{-1}$ and $k \approx 2000$ m$^{-1}$) and two long-scale modes ($k \approx 300$ m$^{-1}$ and $k \approx 600$ m$^{-1}$) with the spectra of the electric field shown in Figs.~\ref{Fig_1D_field_v1} and \ref{Fig_1D_field_v2} for the cases of mobile and immobile ions. The two pronounced short-scale \rv{ion} oscillations ($k \approx 1500$ m$^{-1}$ and $k \approx 2000$ m$^{-1}$) differ \rv{significantly} from the short-scale longitudinal electric field oscillations shown in the lower panel of Fig.~\ref{Fig_1D_field_v2}. On the other hand, one of the short-scale ($k \approx 1500$ m$^{-1}$) and one of the long-scale ($k \approx 300$ m$^{-1}$) modes appear to match the $k$-vectors of the modes that we observed in the simulation with immobile ions.

\begin{figure*} 
\centering
\includegraphics[width=1.5\columnwidth]{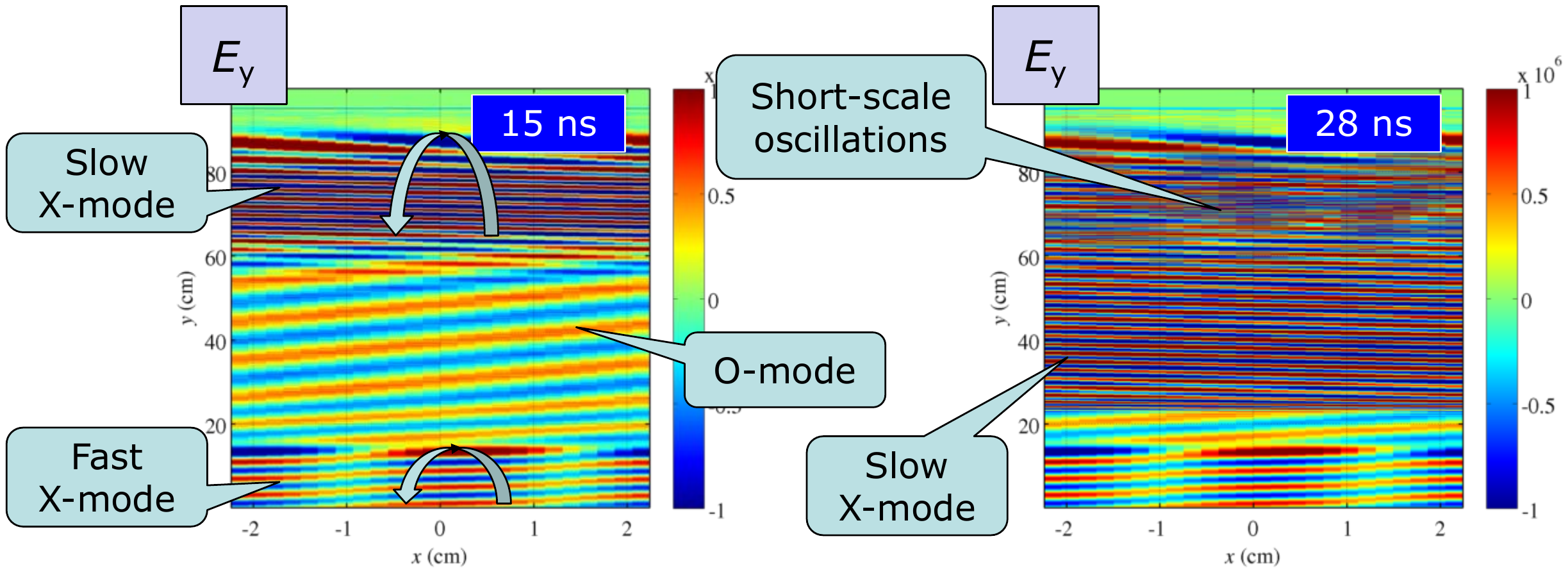}
\caption{Snapshots of the electric field transverse to the confining magnetic field from a 2D PIC simulation of the O-X-B mode conversion in the nonlinear regime with \textit{mobile} ions.}
\label{Fig4}
\end{figure*}

The main reason for the apparent disagreement between the ion spectrum (lower panel of Fig.~\ref{Fig_slow}) and the electric field spectra in Figs.~\ref{Fig_1D_field_v1} and \ref{Fig_1D_field_v2} is that the ion oscillations are low-frequency oscillations, whereas the electric field spectra do not distinguish between low-frequency oscillations and the oscillations at the frequency of the incoming wave. In order to examine low-frequency electric field oscillations, we have performed an additional simulation in which time-averaged electric fields were separately calculated by continuously averaging both components of the field over two cycles of the incoming wave ($f = 10$ GHz). The result is shown in the upper panel of Fig.~\ref{Fig_slow}. The time instant is the same as that for the plot of the ion oscillations in the lower panel of the same figure. The peaks corresponding to the oscillations of the ``slow'' electric field match those in the ion spectrum, which enabled us to identify the corresponding modes.

\begin{figure*} [tb]
\centering
\includegraphics[width=1.5\columnwidth]{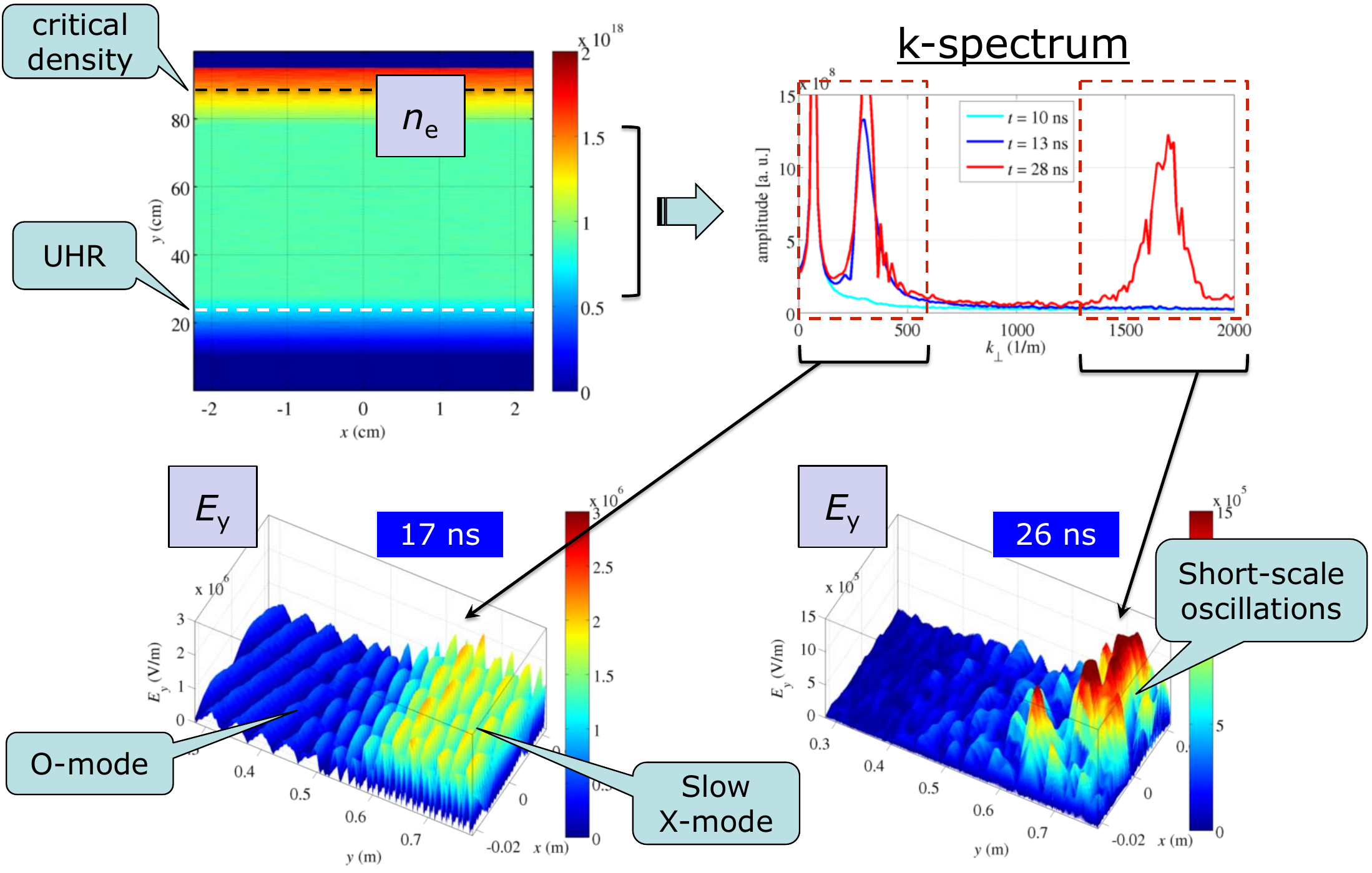}
\caption{2D PIC simulation of the O-X-B mode conversion in the nonlinear regime with \textit{mobile} ions: upper left -- simulation setup; upper right -- $k$-spectrum of the longitudinal electric field $E_y$ at 10 ns, 13, and 28 ns into the simulation in the flat density region. The two lower panels show the long-scale (left) and the short-scale (right) contributions to $E_y$ at 17 ns and at 26 ns. The corresponding ranges of $k_{\perp}$ are marked with dashed boxes in the upper-right panel and connected to the corresponding lower panels with arrows.} \label{Fig5}
\end{figure*}

It must be pointed out that the units used to normalize the vertical scale in Figs.~\ref{Fig_1D_field_v1} and \ref{Fig_1D_field_v2} and in the upper panel of Fig.~\ref{Fig_slow} are the same. The height of the peaks in the slow electric field spectrum is almost an order of magnitude smaller than the height of the peaks in Figs.~\ref{Fig_1D_field_v1} and \ref{Fig_1D_field_v2}. It is therefore not surprising that the slow electric field oscillations are not easily visible without the time-averaging. The bottom line is that the ion dynamics in the flat density region that arises without the converted EBW is coupled to four ``slow'' modes: two short-scale modes and two long-scale modes. 


\section{O-X-B conversion in a nonlinear regime} \label{Sec-OXB}

In order to determine whether the excitation of the short-scale electric fields is a robust feature, we have also carried out a 2D PIC simulation. We access the nonlinear regime by setting the amplitude of the incoming wave, injected at an angle $\theta = 40^{\circ}$ to the density gradient, at $8 \times 10^5$ V/m. The wave frequency is $f = 10$ GHz and its electric field is polarized in the plane of the simulation, i.e. it has both $x$ and $y$-compoments. We use a confining magnetic field, $B_0 =  0.25$ T, directed along the $x$-axis. The size of the computational domain is 1 m ($1.4 \times 10^4$ cells) along the $y$-axis and $c / f \sin \theta$ (25 cells) along the $x$-axis.

The initial electron density profile is given by
\begin{equation}
\left. n_e \right/ n_{crit} =
   \begin{cases}
      0,  & \text{for $y <  y_{b1}$;}\\
      4(y-y_{b1})/l,  & \text{for $y_{b1} \leq y <  y_L$;}\\
      0.72,  & \text{for $y_L \leq y \leq  y_R$;}\\
      4 (y - y_2)/l,  & \text{for $y_R < y <  y_{b2}$;}\\
      0, & \text{for $y \geq y_{b2}$,}
   \end{cases} \label{n_O-X-B_sim}
\end{equation}
where $n_{crit} \approx 1.24 \times 10^{18}$ m$^{-3}$ is the critical density for the \rv{chosen} frequency. We have introduced the following quantities to parameterize the multi-gradient density profile: $y_L = 0.28$ m, $y_R = 0.78$ m, $y_{b1} = 0.1$ m, $y_{b2} = 0.95$ m, $y_1 = 0.1$ m, $y_2 = 0.6$ m, and $l = 1$ m. The electron population is initialized using 400 macro-particles per cell. The initial electron temperature is set at $T_e = 950$ eV. We also initialize cold deuterium ions using 400 macro-particles per cell with the ion \rv{number} density equal to $n_e$.

Figure~\ref{Fig4} shows snapshots of the $E_y$ component of the electric field (the component directed along the density gradient). The 15 ns snapshot captures the initial stage of an O-mode conversion into a slow X-mode. At 28 ns, the slow X-mode has almost reached the UHR layer where the conversion into an EBW takes place in the linear regime. However, short-scale oscillations have already been excited at this stage on the opposite side of the flat density region ($x  > 60$ cm). This behavior is similar to what is observed in the 1D simulation in the nonlinear regime (see Fig.~\ref{Fig3}).

In order to better examine the evolution of the field structure, we have computed the $k$-spectrum of the longitudinal electric field in the flat density region at 10 ns, 13 ns, and 28 ns (see the upper-right panel of Fig.~\ref{Fig5}).  At 10 ns, there is only an O-mode in the flat density region. At 13 ns, a part of the O-mode has already been converted into a slow X-mode that shows up as a peak at around $k_{\perp} \approx 300$ 1/m. At 28 ns, short-scale oscillations also appear in the spectrum of the longitudinal field.  

The lower two panels in Fig.~\ref{Fig5} show the field structure corresponding to the long-scale and short-scale parts of the $k_{\perp}$-spectrum. The left panel presents an expected result with counter-propagating O-mode and slow X-mode. The right panel reveals that the short scale fields are excited not at the UHR, but on the opposite side of the flat density region. Moreover, the short-scale oscillations grow on a \rv{timescale} shorter than the time required for the O-X-B conversion to take place. They appear in the spectrum before the slow X-mode even reaches the conversion layer located at the UHR. This result is consistent with the excitation of short-scale oscillations in the one-dimensional simulation with mobile ions, which confirms the robustness of the observed phenomenon.

We have performed an additional 2D PIC simulation for the same setup, but this time initializing hot ions  ($T_i = 950$ eV). The short-scale oscillations are again excited in the flat density region before the slow X-mode even reaches the conversion layer located at the UHR. Additional runs with heavier cold ions show that the amplitude of the short-scale oscillations reduces as we increase the ion mass. This is consistent with our conclusion that it is the ion mobility that enables the excitation of the short-scale oscillations in the plasma.


\section{Theoretical interpretation} \label{Sec-Theory}

We interpret the appearance of the small-scale oscillations that were described above as an observation of the X-wave instability with respect to scattering into an EBW and a lower-hybrid wave (LHW). This instability was previously discussed in a number of papers; for example, see Refs.~\cite{Gusakov2007, McDermott1982, Porkolab1982, Lin1981}. However, to our knowledge, we are first to report an observation of this instability in PIC simulations and explore it in detail.

In order to interpret the structure of the observed spatial spectra in the context of this instability in simplest terms, we focus on the explanation of the 2D simulations. {\rv T}he scattering can be understood as a three-wave interaction satisfying two scalar resonance conditions:
\begin{gather}\label{eq:rc}
\omega\wX \approx \omega\wB + \omega\wLH, \quad 
k_x\wX \approx k_x\wB + k_x\wLH.
\end{gather}
Here $\smash{\omega\wQ}$ are frequencies and $k\wQ_x$ are projections of the corresponding wave vectors on the axis of propagation~$x$. One can expect $\smash{\omega\wLH \approx \omega_{\rm LH} \ll \omega\wX \sim \omega\wB \sim \omega_{\rm UH}}$, where $\omega_{\rm LH}$ and $\omega_{\rm UH}$ are the lower- and upper-hybrid frequency, respectively. Thus EBWs must be excited at the frequency $\omega\wB \approx \omega\wX - \omega\wLH$. The corresponding wave numbers found from the linear dispersion relation are $\smash{k\wB_x} \approx \pm 1750\,\text{m}^{-1}$. Note that the part of the EBW that is produced through \textit{linear} conversion (without an LHW \rv{being} involved) has frequency $\omega\wB = \omega\wX$, so its $\smash{k\wB_x}$ must be slightly lower, namely, $k\wB \approx 1600\,\text{m}^{-3}$\rv{: t}his difference can \rv{indeed} be seen in Fig.~3(d).

As determined by the cold linear dispersion relation, the incident and reflected X waves have wave numbers $\smash{k_x\wX \approx \pm 300\,\text{m}^{-1}}$, in agreement with simulations. Hence, Eq.~(\ref{eq:rc}) predicts two possible values for $\smash{k_x\wLH}$, namely, $\smash{\pm 1450\,\text{m}^{-1}}$ and $\smash{\pm 2050\,\text{m}^{-1}}$. This is in agreement with Fig.~7, which shows pronounced peaks in the electric field and ion momentum spectra at these expected locations. (Positive and negative $k$ are not distinguished in our figures.) We also point out that, under the specified parameters, $2\pi/\omega_{\rm LH} \approx 7.9\,\text{ns}$. The temporal oscillations in Fig.~6(c) have approximately this period, so they can \rv{indeed} be safely identified as LH oscillations. The peak at $\smash{|k| \sim 600\,\text{m}^{-1}}$ is due to the beating of the LHWs plus the contribution of the X-wave\rv{'s} second harmonic. The peak at $\smash{|k| \sim 300\,\text{m}^{-1}}$ corresponds to the incident and reflected X waves. (Ion \textit{velocities} cannot be perturbed by the X waves because such waves have frequencies \rv{that are too high}; however, ion \textit{momenta} can be.) The reason why the X-wave fields show up on a figure for time-averaged quantities is that the averaging is performed over the period of the unperturbed incident wave. The actual X waves have non-stationary envelopes evolving at rates $\sim \omega_{\rm LH}$, so averaging of these fields does not eliminate them entirely.

A detailed analytical study of the instability discussed here is somewhat lengthy, so we report it in a separate paper~\cite{Dodin2017}. Here we present the final answer only. In the regime of interest, the instability rate is expected to be
\begin{gather}\label{eq:gamma}
\gamma \approx \omega_{\rm LH}\,\frac{\omega_{pe}\omega_{pi}}{|\Omega_e\Omega_i|}
\sqrt{\frac{\omega_{\rm LH}}{\omega_{\rm UH}}}\,\frac{\big|k\wLH E\wX_x\big|}{16 \pi e n_e}.
\end{gather}
Here, $\omega_{pe}$ and $\omega_{pi}$ are the electron and ion plasma frequencies \rv{respectively}, $\Omega_e$ and $\Omega_i$ are the electron and ion cyclotron frequencies, $e$ is the electron charge, $n_e$ is the unperturbed electron density, and $\smash{E\wX_x}$ is the $x$ component of the X-wave electric field. Note that, for \rv{sufficiently} small temperature $T$, the theory predicts $\gamma$ to have no dependence on $T$, while the dependence on the ion mass remains. This is precisely what is observed in our simulations. More specifically, for our parameters, \Eq{eq:gamma} gives $\gamma^{-1} \approx 7.6\,\text{ns}$. This is in reasonable agreement with Fig.~6(c), especially considering that the figure illustrates the already nonlinear regime and that \Eq{eq:gamma} entirely ignores thermal corrections. The fact that the theory agrees with the simulation results serves as an indicator that the underlying physical process is identified correctly: \rv{namely that} the small-scale oscillations seen in simulations are a result of the X-wave instability with respect to scattering into an EBW and a LHW.


\section{Summary} \label{Sec-Summary}

We have simulated X-B and O-X-B mode conversion in a nonlinear regime where the driving wave appreciably distorts electron thermal motion. We find that the ion mobility has a profound effect on the field structure in this regime. High amplitude short-scale oscillations of the longitudinal electric field are observed in the region below the high-density cut-off prior to the arrival of the EBW. We have additionally performed an extensive parameter scan by varying the spatial resolution, the ion mass, and the ion temperature. This scan shows the robustness of the observed effect. 

We identify this effect as the instability of the X wave with respect to resonant scattering into an EBW and a lower-hybrid wave. We calculate the instability rate analytically and find this basic theory to be in reasonable agreement with our simulation results.

\rv{More generally, the methodology introduced in this paper provides the framework for an accurate quantitative assessment of the feasibility of heating and current-drive in next-generation spherical tokamaks by megawatt-level beams.}


\section{Acknowledgments}
This work was funded in part by the US Department of Energy under grants DE-FG02-04ER54742 and DE-AC02-09CH11466, the University of York, the UK EPSRC under grant EP/G003955, and the European Communities under the contract of Association between EURATOM and CCFE. Simulations were performed using the EPOCH code (developed under UK EPSRC grants EP/G054940, EP/G055165 and EP/G056803) using HPC resources provided by the Texas Advanced Computing Center at The University of Texas and using the HELIOS supercomputer system at Computational Simulation Centre of International Fusion Energy Research Centre (IFERC-CSC), Aomori, Japan, under the Broader Approach collaboration between Euratom and Japan, implemented by Fusion for Energy and JAEA.

\end{document}